\documentstyle[twoside,fleqn]{article}
\makeatletter
\def\fileversion{v2.6}
\def\filedate{24 November 1993}

\newdimen\@bls                    
\@bls=\baselineskip               
\advance\@bls -1ex                
\newdimen\@eps                    %
\def\section{\@startsection{section}{1}{\z@}
  {1.5\@bls plus 0.5\@bls}{1\@bls}{\normalsize\bf}}
\def\subsection{\@startsection{subsection}{2}{\z@}
  {1\@bls plus 0.25\@bls}{\@eps}{\normalsize\bf}}
\def\subsubsection{\@startsection{subsubsection}{3}{\z@}
  {1\@bls plus 0.25\@bls}{\@eps}{\normalsize\bf}}
\def\paragraph{\@startsection{paragraph}{4}{\parindent}
  {1\@bls plus 0.25\@bls}{0.5em}{\normalsize\bf}}
\def\subparagraph{\@startsection{subparagraph}{4}{\parindent}
  {1\@bls plus 0.25\@bls}{0.5em}{\normalsize\bf}}

\def\@sect#1#2#3#4#5#6[#7]#8{\ifnum #2>\c@secnumdepth
  \def\@svsec{}\else
  \refstepcounter{#1}\edef\@svsec{\csname the#1\endcsname.\hskip0.5em}\fi
  \@tempskipa #5\relax
  \ifdim \@tempskipa>\z@
    \begingroup
      #6\relax
      \@hangfrom{\hskip #3\relax\@svsec}{\interlinepenalty \@M #8\par}%
    \endgroup
    \csname #1mark\endcsname{#7}\addcontentsline
      {toc}{#1}{\ifnum #2>\c@secnumdepth \else
        \protect\numberline{\csname the#1\endcsname}\fi #7}%
  \else
    \def\@svsechd{#6\hskip #3\@svsec #8\csname #1mark\endcsname
      {#7}\addcontentsline{toc}{#1}{\ifnum #2>\c@secnumdepth \else
        \protect\numberline{\csname the#1\endcsname}\fi #7}}%
  \fi \@xsect{#5}}

\long\def\@makefigurecaption#1#2{\vskip 10mm #1. #2\par}

\long\def\@maketablecaption#1#2{\hbox to \hsize{\parbox[t]{\hsize}
  {#1 \\ #2}}\vskip 0.3ex}

\def\fnum@figure{Figure \thefigure}
\def\figure{\let\@makecaption\@makefigurecaption \@float{figure}}
\@namedef{figure*}{\let\@makecaption\@makefigurecaption \@dblfloat{figure}}

\def\table{\let\@makecaption\@maketablecaption \@float{table}}
\@namedef{table*}{\let\@makecaption\@maketablecaption \@dblfloat{table}}

\textfloatsep=\floatsep           
\intextsep=\floatsep              

\long\def\@makefntext#1{\parindent 1em\noindent\hbox{${}^{\@thefnmark}$}#1}

\def\maketitle{\begingroup        
    \def\thefootnote{\fnsymbol{footnote}}%
    \newpage \global\@topnum\z@
    \@maketitle \@thanks
  \endgroup
  \let\maketitle\relax \let\@maketitle\relax
  \gdef\@thanks{}\let\thanks\relax
  \gdef\@address{}\gdef\@author{}\gdef\@title{}\let\address\relax}

\def\justify@on{\let\\=\@normalcr
  \leftskip\z@ \@rightskip\z@ \rightskip\@rightskip}

\newbox\fm@box                    

\def\@maketitle{
  \global\setbox\fm@box=\vbox\bgroup
    \vskip 8mm                    
    \raggedright                  
    \hyphenpenalty\@M             
    {\Large \@title \par}         
    \vskip\@bls                   
    {\normalsize                  
     \@author \par}               
    \vskip\@bls                   
    \@address                     
  \egroup
  \twocolumn[
    \unvbox\fm@box                
    \vskip\@bls                   
    \unvbox\abstract@box          
    \vskip 2pc]}                  

\newcounter{address}
\def\theaddress{\alph{address}}
\def\@makeadmark#1{\hbox{$^{\rm #1}$}}

\def\address#1{\addressmark\begingroup
  \xdef\@tempa{\theaddress}\let\\=\relax
  \def\protect{\noexpand\protect\noexpand}\xdef\@address{\@address
  \protect\addresstext{\@tempa}{#1}}\endgroup}
\def\@address{}

\def\addressmark{\stepcounter{address}%
  \xdef\@tempb{\theaddress}\@makeadmark{\@tempb}}

\def\addresstext#1#2{\leavevmode \begingroup
  \raggedright \hyphenpenalty\@M \@makeadmark{#1}#2\par \endgroup
  \vskip\@bls}

\newbox\abstract@box              

\def\abstract{%
  \global\setbox\abstract@box=\vbox\bgroup
  \small\rm
  \ignorespaces}
\def\endabstract{\par \egroup}

\def\thebibliography#1{\section*{REFERENCES}\list{\arabic{enumi}.}
  {\settowidth\labelwidth{#1.}\leftmargin=1.67em
   \labelsep\leftmargin \advance\labelsep-\labelwidth
   \itemsep\z@ \parsep\z@
   \usecounter{enumi}}\def\makelabel##1{\rlap{##1}\hss}%
   \def\newblock{\hskip 0.11em plus 0.33em minus -0.07em}
   \sloppy \clubpenalty=4000 \widowpenalty=4000 \sfcode`\.=1000\relax}

\newcount\@tempcntc
\def\@citex[#1]#2{\if@filesw\immediate\write\@auxout{\string\citation{#2}}\fi
  \@tempcnta\z@\@tempcntb\m@ne\def\@citea{}\@cite{\@for\@citeb:=#2\do
    {\@ifundefined
       {b@\@citeb}{\@citeo\@tempcntb\m@ne\@citea
        \def\@citea{,\penalty\@m\ }{\bf ?}\@warning
       {Citation `\@citeb' on page \thepage \space undefined}}%
    {\setbox\z@\hbox{\global\@tempcntc0\csname b@\@citeb\endcsname\relax}%
     \ifnum\@tempcntc=\z@ \@citeo\@tempcntb\m@ne
       \@citea\def\@citea{,\penalty\@m}
       \hbox{\csname b@\@citeb\endcsname}%
     \else
      \advance\@tempcntb\@ne
      \ifnum\@tempcntb=\@tempcntc
      \else\advance\@tempcntb\m@ne\@citeo
      \@tempcnta\@tempcntc\@tempcntb\@tempcntc\fi\fi}}\@citeo}{#1}}

\def\@citeo{\ifnum\@tempcnta>\@tempcntb\else\@citea
  \def\@citea{,\penalty\@m}%
  \ifnum\@tempcnta=\@tempcntb\the\@tempcnta\else
   {\advance\@tempcnta\@ne\ifnum\@tempcnta=\@tempcntb \else
\def\@citea{--}\fi
    \advance\@tempcnta\m@ne\the\@tempcnta\@citea\the\@tempcntb}\fi\fi}

\def\ps@crcplain{\let\@mkboth\@gobbletwo
     \def\@oddhead{\reset@font{\sl\rightmark}\hfil \rm\thepage}%
     \def\@evenhead{\reset@font\rm \thepage\hfil\sl\leftmark}%
     \let\@oddfoot\@empty
     \let\@evenfoot\@oddfoot}

\ps@crcplain                    

\newcommand{\AmS}{{\protect\the\textfont2
  A\kern-.1667em\lower.5ex\hbox{M}\kern-.125emS}}

\makeatother

\title{Casimir energy of dielectric systems}

\author{{Valery N. Marachevsky }
\address{Department of Theoretical Physics, St.Petersburg State
University, 198504 St.Petersburg, Russia}
\thanks{E-mail: root@VM1485.spb.edu, maraval@mail.ru}}

\begin{document}
\newcommand{\om}{\bigl|\omega\bigr|\sqrt{\varepsilon}}
\newcommand{\omm}{\bigl|\omega\bigr|}

\begin{abstract}
\qquad
A new formula for the Casimir energy of a dispersive
dilute dielectric ball is discussed. The formula
for the Casimir energy of a polarizable particle
situated in a perfectly conducting wedge-shaped cavity is derived by
a path-integral coordinate space method in quantum field theory.

\end{abstract}

\maketitle


\section{Introduction}
This paper is devoted to study of Casimir effect \cite{C1} in dielectrics
at zero temperature in the framework of quantum field theory.
The presentation of the subject is based on the results and proofs derived by the author.

The paper is organized as follows.
In section $2$ a new formula for the Casimir energy
of a dispersive dilute dielectric ball \cite{Mar10} is discussed.
In section $3$ a path-integral method is used to justify the formalism
that was originally developed by Lifshitz
in the framework of statistical physics \cite{Lif1,Lifshitz}.
This method was applied in Ref. \cite{Mar3} to derive
for the first time a formula
for the Casimir energy of a polarizable
particle situated in a perfectly conducting
wedge-shaped cavity.
These examples illustrate two different regimes:
a dilute connected dielectric with pairwise dipole-dipole interaction
between atoms, and the system of two disjoint dielectrics,
one of which is not a dilute one,
where many-body effects of non-pairwise interaction are important.

We put $\hbar=c=1$ and use rationalized Gaussian units where
the polarizability of atoms $\alpha(i\omega)$ is defined via
$\varepsilon(i\omega) - 1 = 4\pi \rho\,\alpha(i\omega)$,
$\rho$ is a number density of atoms,
$\omega_0$ is a characteristic absorption frequency of materials,
$\lambda$ is an average minimum distance between atoms of a dielectric,
Casimir--Polder  potential
is defined by $-23\,\alpha_1(0)\alpha_2(0)/4\pi r^7$.

\section{Casimir energy of a dispersive dilute dielectric ball}
We study a dielectric nonmagnetic  ball of the radius $a$ and permittivity
$\varepsilon$, surrounded  by  a vacuum.  The ball is dilute, i.e. all
final expressions  are obtained under the assumption $\varepsilon - 1
\ll 1$ in the order $(\varepsilon(i\omega) - 1)^2$, the lowest order that
yields the  energy of interaction between atoms of the ball.

The study of the Casimir energy of a nonmagnetic
dielectric ball remains one of the main  problems
in the theory of Casimir effect.
Cut-off dependent terms arised in every
macroscopic approach to the problem,
so it was possible to extract correctly
only the large distance contribution to the energy
of a dilute dielectric ball by use of macroscopic methods
(see Refs. \cite{Mar1}, \cite{Vas} and Appendix in Ref.\cite{Mar}).
The regularization of ill-defined expressions
remains the main problem of various macroscopic approaches
to connected dielectrics. Usually the Casimir energy
of a  disjoint macroscopic system
(two dispersive dielectric parallel plates is
a classic example by Lifshitz) depends only on the
distance between macroscopic bodies
and dispersion of dielectrics \cite{Lifshitz}.
On the other hand, it was argued in Refs. \cite{Mar}, \cite{Barton} and \cite{Mar10}
that for a dilute connected dielectric the Casimir energy is
equal to the energy of dipole-dipole pairwise interactions
of all atoms constituting the dielectric
and thus should also depend on an average minimum distance between
atoms of a dielectric $\lambda$.
For a dilute dispersive dielectric ball with an arbitrary
frequency dependent dielectric permittivity
the Casimir energy was first derived in Ref. \cite{Mar10}:
\begin{eqnarray}
\lefteqn{E = - \rho^2 \frac{\pi}{48} \int_{0}^{+\infty} d\omega
 \alpha^2(i\omega) } \nonumber\\
\lefteqn{ \Bigl( \frac{a^3}{\lambda^3} e^{-2\omega\lambda} (128 + 256 \omega\lambda +
 128 \omega^2\lambda^2  + 64 \omega^3\lambda^3 )  } \nonumber\\
\lefteqn{ - \frac{a^2}{\lambda^2} \bigl(e^{-2 \omega\lambda}(144 + 288\omega\lambda +
 120\omega^2\lambda^2 + 48\omega^3\lambda^3 )  } \nonumber\\
\lefteqn{-96\omega^2\lambda^2
 E_1(2\omega\lambda) \bigr) } \nonumber \\
+ \lefteqn{ \bigl(e^{-2\omega\lambda}(41 +
 34\omega\lambda + 14\omega^2\lambda^2 + 4\omega^3\lambda^3 ) } \label{tt3} \\
\lefteqn{+24 E_1(2\omega\lambda)\bigr) +} \nonumber\\
\lefteqn{ \bigl(e^{-4\omega a} (-21 + 12\omega a) -
 E_1(4\omega a) (24 + 96 \omega^2 a^2) \bigr)
 \Bigr).} \nonumber \end{eqnarray}
Here $E_1(x)=\int_{1}^{+\infty} e^{-tx}/t \, dt$.

This energy is finite and physical only when a finite separation
between atoms $\lambda$ is taken into account. In the hypothetical
non-physical limit $\lambda \to 0$ the leading term in (\ref{tt3})
($V$ is a ball volume) is
\begin{equation}
- \rho^2 \frac{2 V }{\lambda^3} \int_{0}^{+\infty} d\omega \,
 \alpha^2 (i\omega), \label{tt4}
\end{equation}
so in the limit $\lambda \to 0$ the Casimir energy is divergent
for every model of atomic polarizability $\alpha(i\omega)$.

The formula (\ref{tt3}) finally solves the problem of the Casimir
energy for a dilute dielectric ball. The interest to this topic
strongly arised after the series of articles by Schwinger
\cite{Schwinger} where he tried to treat the sonoluminescence of
bubbles in water \cite{Sono} as a dynamical Casimir effect.
Schwinger suggested that the first order terms $\sim (\varepsilon
-1) V$ should yield the main contribution to the energy of the
ball. One possible argument why the first
order terms have no influence on physics is the condition of the
conservation of atoms constituting the ball:
\begin{equation}
(\varepsilon - 1) V = {\rm const} .
\end{equation}
It follows from this condition that the first order terms
do not change their value during the ball
collapse or expansion, so they can be subtracted from the energy.

After publication of the articles \cite{Mar1}
it was generally believed that
the Casimir energy of a dilute dielectric
ball is equal to (this expression was first derived in \cite{M1})
\begin{equation}
E_{ld} = \frac{23}{1536\pi a} (\varepsilon - 1)^2 , \label{ad1}
\end{equation}
which is only a non-dispersive limit of the last line
of our new full expression for the energy (\ref{tt3}).
To check this, we write the leading contribution
from the last line of (\ref{tt3}) as (it comes from frequencies
$\omega \ll \omega_0$, $\omega_0$ is a characteristic absorption
frequency of materials, $\omega_0 a \gg 1$, so it is possible to
use the static polarizability $\alpha(0)$ in the leading
approximation to the last line of (\ref{tt3}))
\begin{eqnarray}
\lefteqn{-\rho^2 \alpha^2(0) \frac{\pi}{48} \int_{0}^{+\infty} d\omega  \,
  \bigl(e^{-4\omega a} (-21 + 12\omega a) } \nonumber \\
\lefteqn{-E_1(4\omega a) (24 + 96 \omega^2 a^2) \bigr) } \nonumber \\
\lefteqn{ = \rho^2 \alpha^2(0) \frac{23}{96} \frac{\pi}{a} =
  \frac{23}{1536\pi a} (\varepsilon - 1)^2 = E_{ld}. } \label{tt5}
\end{eqnarray}
The new full energy expression
contains additional terms: volume
and surface contributions to the energy,  as well as the terms
which do not depend on the ball radius $a$.

The term (\ref{ad1})
can be called a large distance contribution to the  Casimir energy of the
ball. However, it is impossible to separate large distances
between atoms of a dielectric ball from short distances between atoms
of the ball in any possible experiment. This is why only the use of a
dipole-dipole potential valid for all existing distances
between  atoms of the ball (which are all greater than $\lambda$)
is physically reasonable for the calculation of Casimir energies
of dilute bodies.
To understand this, consider the line of
the following examples.
Casimir energy of two neutral atoms coincides with the energy of a
dipole-dipole interaction of these atoms. When an atom is located
outside a dielectric of an arbitrary form,
then in a dilute approximation the Casimir energy is
equal to the sum of dipole-dipole interactions between this atom
and atoms of the dielectric \cite{Mar}.
For two parallel dielectric slabs
it is known that the Casimir energy in a dilute approximation
is equal to the sum of pairwise dipole-dipole
interactions of atoms constituting the slabs. For a dilute dielectric
ball the sum of pairwise dipole-dipole interactions of atoms
constituting the ball is given by Eq.(\ref{tt3}), not Eq.(\ref{ad1}).

There are several important  differences between our microscopic
calculation \cite{Mar10} and microscopic calculation in \cite{M1}
where the term (\ref{ad1}) was first derived.
In the article \cite{M1}  the Casimir-Polder potential,
which is a large distance limit of a dipole-dipole potential,
valid only for distances $r\gg 1/\omega_0$, was
used in the calculation of the ball Casimir energy.
The calculation in Ref. \cite{M1} was non-dispersive from the
outset and thus could not yield the contribution to the energy from
short distances ($r < 1/\omega_0$) between atoms of the
ball. Moreover, the use of a dimensional regularization in
Ref. \cite{M1} concealed the divergences which would appear in the
energy expression from the integration over short distances
between atoms in the 3-dimensional ball since the minimum distance
between atoms of the ball was not introduced in Ref. \cite{M1}.

Needless to say that the term (\ref{ad1}) itself was really
important for development of the theory of Casimir effect in
connected dielectrics since this term has been derived via
different techniques \cite{M1,Mar1}.
However, using these approaches one can extract correctly only the large distance
contribution to the Casimir energy of the ball, e.g.$E_{ld}$. This large
distance contribution $E_{ld}$ was found to be the
same when summing up the Casimir--Polder
potential between atoms of the ball \cite{M1} and when
the Casimir energy was derived by
field-theoretic calculations \cite{Mar1} - so the
equivalence of large distance parts of the Casimir energy for a
dilute dielectric ball derived
by microscopic and macroscopic approaches was proved.

It is important
to stress that so far macroscopic  methods did not yield
satisfactorily short distance contributions to the Casimir energy
of connected dielectrics. The reason is simple: these methods
were developed for {\it disjoint}, not connected dielectrics,
and application of these methods to connected
dielectrics without any changes inevitably leads
to different types of divergences in {\it every}
field-theoretic calculation of the Casimir energy. These divergences are
reminiscents of the ill-defined short distance structure of the
theory.

It was generally believed that Casimir surface force is repulsive
before the appearance of our papers \cite{Mar,Mar10}.
So it is natural to give here a proof that Casimir surface force on a
dilute dielectric ball is attractive.

It is convenient to define
$N\equiv a/\lambda, p\equiv\omega\lambda$. Then Eq.(\ref{tt3})
can be rewritten in a general form
\begin{equation}
E = - \frac{\rho^2}{\lambda} \int_{0}^{+\infty} dp \, \alpha^2\Bigl(i
\frac{p}{\lambda}\Bigr) f(N,p) .   \label{tt6}
\end{equation}
The function $f(N,p)>0$ for $N>1/2, p>0$. The ball expands or
collapses homogeneously, so
\begin{equation}
N={\rm const}. \label{tt7}
\end{equation}
Conservation of atoms inside
the ball imposes the condition
\begin{equation}
\rho \, \frac{4\pi a^3}{3} = const. \label{tt8}
\end{equation}
It is convenient to use  Kramers--Kronig relations in the form
\begin{equation}
\alpha(i\omega) = \int_{0}^{+\infty} dx\, \frac{x g(x)}{x^2+\omega^2}, \label{tt9}
\end{equation}
where the condition $g(x)>0$ always holds.
Using (\ref{tt6}), (\ref{tt7}), (\ref{tt8}), (\ref{tt9}), Casimir
force on a unit surface is equal to
\begin{eqnarray}
\lefteqn{ F = -\frac{1}{4\pi a^2} \frac{\partial E}{\partial a} =
-\frac{\rho^2}{4\pi a^3}\int_{0}^{+\infty} d\omega \int_{0}^{+\infty}
dx} \nonumber \\ \lefteqn{ \frac{x (7x^2 + 3 \omega^2) g(x)}{(x^2+\omega^2)^2}\,
\alpha(i\omega) f(N,\omega \lambda)  < 0. } \label{tt10}
\end{eqnarray}
$F<0$ because all functions inside integrals are positive. Casimir
surface force is attractive for every model of atomic
polarizability consistent with general causal requirements.

\section{Casimir energy of a polarizable particle in a perfectly conducting wedge}
The one-loop effective action,
which is the result of the integration over quantum fluctuations of the electromagnetic field,
has the form
\begin{equation}
W = \frac{1}{2} {\rm Tr} \ln L_{pp} (\varepsilon, t, {\bf x}),  \label{q1}
\end{equation}
where
\begin{equation}
L_{jm} (\varepsilon, \omega, {\bf r}) =
\Bigl[ \varepsilon (i |\omega|, {\bf r}) \, \omega^2 \delta_{jm}  +
{\rm  rot}_{jl}{\rm rot}_{lm} \Bigr]  \, .  \label{p21}
\end{equation}
Electromagnetic field propagator
$D_{mk}(\varepsilon, \omega, {\bf r},{\bf r}^\prime) $
in a gauge $A_0=0$ satisfies the equations
\begin{equation}
L_{jm} (\varepsilon, \omega, {\bf r})
D_{mk}(\varepsilon, \omega, {\bf r},{\bf r}^\prime)
 =  \delta ({\bf r} - {\bf r}^\prime)
\delta_{jk} \, . \label{p22}
\end{equation}

Imagine that the space was empty at first, with no dielectric in it.
Imagine then that the particle with an atomic polarizability
$\alpha(i|\omega|)$ is inserted at the point ${\bf x}={\bf r^\prime}$ so that
$\delta \varepsilon(i|\omega|, {\bf x}) = 4\pi \alpha(i|\omega|)\,
\delta^3({\bf x}-{\bf r^\prime}) $.
The change in the ground state energy for a small $\alpha(i|\omega|)$ is equal to
\begin{equation}
\delta E_1 =  \int_{-\infty}^{+\infty}
d\omega \,  \alpha(i|\omega|) \, \omega^2
D_{pp}(\varepsilon=1, \omega, {\bf r^\prime}, {\bf r^\prime}) \, .
\end{equation}
This change is divergent. However, there are no other particles around and
thus this is not the energy of interaction that can be measured.

Imagine now that the same particle is inserted in the neighbourhood
of a dielectric body with an arbitrary permittivity
$\varepsilon(i|\omega|, {\bf r_2})$.
The change in the ground state energy in this case is given by
\begin{equation}
\delta E_2 =  \int_{-\infty}^{+\infty}
d\omega \,  \alpha(i|\omega|) \, \omega^2
D_{pp}(\varepsilon, \omega, {\bf r^\prime}, {\bf r^\prime}) \, ,
\end{equation}
but this is not the answer for the energy of interaction.
The energy responsible for interaction of a particle and a
dielectric is finite and equal to $\delta E = \delta E_2 - \delta E_1$:
\begin{eqnarray}
\delta E &=& \int_{-\infty}^{+\infty}
d\omega \,  \alpha(i|\omega|) \, \omega^2
( D_{pp}(\varepsilon, \omega, {\bf r^{\prime\prime}}, {\bf r^\prime})
 \nonumber \\
&\,& - D_{pp}(\varepsilon =1, \omega, {\bf r^{\prime\prime}}, {\bf r^\prime}))
\Bigr|_{ {\bf r^{\prime\prime}} \to {\bf r^\prime} } . \label{tm2}
\end{eqnarray}

This formula can be used to calculate the Casimir energy of
a polarizable particle with the polarizability $\alpha(i\omega)$
located at the point with cylindrical coordinates ($r, \theta, z$)
in a perfectly conducting wedge-shaped cavity \cite{Mar3}
(the walls of the wedge have coordinates
($r,0,z$) and ($r, \alpha, z$), $0 < \theta < \alpha$).

This system may be described by
the set of equations (\ref{p22}) with $\varepsilon=1$ outside the wedge walls
and perfect boundary conditions imposed on each spatial argument of
$D_{ij} (\varepsilon=1, \omega, {\bf r}, {\bf r^\prime})$
at the wedge walls.

For distances $r\theta, r(\alpha-\theta) \gg \lambda_0 \sim 50$nm
one can neglect dispersion in
an atomic polarizability of the particle and walls of the wedge
and take the limit $\alpha(0)$ from the beginning.
Perfectly conducting walls of the wedge can be considered as the
limiting case of the walls with a constant permittivity
$\varepsilon(0)$ when $\varepsilon(0) \to \infty$.

Casimir energy of a polarizable particle situated in a perfectly
conducting wedge-shaped cavity
was first calculated in Ref. \cite{Mar3}.
It can be derived from (\ref{tm2}) in the form:
\begin{eqnarray}
\varepsilon({\bf r}) &=&
-\frac{\alpha(0)}{4 \pi r^4}
\Biggl[\frac{3}{2}\frac{p^4}{\sin^4 p\theta}
- \frac{p^2(p-1)(p+1)}{\sin^2 p\theta}   \nonumber
\\&\,& -\frac{1}{90} (p-1)(p+1)(p^2 +
11)\Biggr] ,  \label{Z2}
\end{eqnarray}
where $ p=\frac{\pi}{\alpha} $ .


\begin{thebibliography}{99}

\bibitem{C1} H. B. G. Casimir, Proc. K. Ned. Akad. Wet. 51 (1948) 793.

\bibitem{Mar10} V. N. Marachevsky, Mod. Phys. Lett. A 16 (2001) 1007
(HEP-TH 0101062).

\bibitem{Lif1} E. M. Lifshitz, Zh. Eksp. Theor. Fiz. 29 (1955) 94.

\bibitem{Lifshitz}   E. M. Lifshitz and  L. P. Pitaevskii, Statistical
Physics, Part 2 (Course of Theoretical Physics, vol. IX), Moscow,
Nauka, 1978, Chapter 8.

\bibitem{Mar3}  I. Brevik, M. Lygren   and  V. N. Marachevsky,
Ann. Phys. (N.Y.) 267 (1998) 134.

\bibitem{Mar1}
I. Brevik and V. Marachevsky,
Phys. Rev. D 60 (1999) 085006.
I. Brevik, V. N. Marachevsky  and  K. A. Milton, Phys. Rev. Lett. 82 (1999)
3948  (HEP-TH 9810062).
G. Barton, J. Phys. A 32 (1999) 525.

\bibitem{Vas}
M. Bordag, K. Kirsten and D. Vassilevich, Phys. Rev. D 59 (1999)
085011 (HEP-TH 9811015).

\bibitem{Mar} V. N. Marachevsky, Phys.Script. 64 (2001) 205 (HEP-TH 0010214).


\bibitem{Barton}
G. Barton, J. Phys. A 34 (2001) 4083.

\bibitem{Schwinger}  J. Schwinger,   Proc. Natl. Acad. Sci.  USA
90 (1993) 958, 2105, 4505, 7285; 91 (1994) 6473.

\bibitem{Sono} B. P. Barber, R. A. Hiller, L. L\"ofstedt, S. J. Putterman and
K. Weniger,  Phys. Rep. C 281 (1997) 65.
M. A. Margulis, Usp. Fiz. Nauk 170 (2000) 263.

\bibitem{M1}
K. A. Milton and  Y. J. Ng,  Phys. Rev. E  57 (1998) 5504.

\end{thebibliography}
\end{document}